\RequirePackage[2020-02-02]{latexrelease}
\documentclass[twocolumn,preprintnumbers,amsmath,amssymb]{revtex4}
\usepackage{graphicx}
\usepackage{dcolumn}
\usepackage{bm}
\usepackage{color}
\usepackage{physics,dsfont}
\usepackage{subfigure}
\usepackage{ulem}

\begin{document}
	
	\title{Multichromatic Floquet engineering of quantum dissipation}

	\author{Fran\c{c}ois Impens$^1$ and David Gu\'ery-Odelin$^2$}
\affiliation{$^1$ Instituto de F\'{i}sica, Universidade Federal do Rio de Janeiro,  Rio de Janeiro, RJ 21941-972, Brazil
\\
$^2$ Laboratoire Collisions, Agr\'egats, R\'eactivit\'e, FeRMI, Universit\'e de Toulouse, CNRS, UPS, France}

	\begin{abstract}
		The monochromatic driving of a quantum system is a successful technique in quantum simulations, well captured by an effective Hamiltonian approach, and with applications in artificial gauge fields and topological engineering. In this letter, we investigate the modeling of multichromatic Floquet driving for the slow degrees of freedom. Within a well-defined range of parameters, we show that the time coarse-grained dynamics of such a driven closed quantum system is encapsulated in an effective Master equation for the time-averaged density matrix, that evolves under the action of an effective Hamiltonian and tunable Lindblad-type dissipation/quantum gain terms. As an application, we emulate the dissipation induced by phase noise and incoherent emission/absorption processes in the bichromatic driving of a two-level system.
	\end{abstract}

\maketitle

There is currently an intense research effort devoted to the realization of quantum simulators able to reproduce complex quantum dynamics in simpler and controlable setups~\cite{QuantumSimulatorsReview}. In many cases, the quantum systems to be emulated are coupled to an environment, and thus behave as open quantum systems. Such an interaction is usually considered as detrimental. However, a controlled dissipation can be a unique asset for quantum state targetting \cite{RCGG20} such as ground state \cite{PFC18}, pointer state \cite{dqs,ZurekRMP03}, or even excited state \cite{RPD14}, and opens many perspective for many-body  quantum simulation \cite{RMK20}. 

The emulation of quantum dissipation is therefore an important step in the roadmap to accurate quantum simulators. Several mechanisms have been used to produce dissipation in a quantum setup. It includes the driving of two interacting quantum subsystems - one of them acting as a bath on the other~\cite{Diehl08,Diehl10}, the use of atom losses for studying loss cooling \cite{Ref15,Ref16}, the Zeno effect \cite{Ref9,Ref10,Ref11,Ref12,Ref13}, the bi-stability of atom transport \cite{Ref14}, the control of decoherence effects~\cite{FloquetDecoherence15}  and the investigation of many-body phase transition with dissipative phenomena \cite{Ref17} to name a few. In this letter, we detail an alternative strategy relying on multichromatic Floquet driving to emulate quantum dissipation while keeping the system conservative.  

Periodic Floquet-driven quantum systems have become instrumental to emulate novel interactions, quantum states of matter or artificial gauge fields~\cite{Rahav03,Goldman14,FloquetReview1,Goldman15,Eckardt15,Itin15,Mikami16,FloquetReview2,FloquetTopologicalInsulators,FloquetReview3,FloquetReview4,FloquetReview5}.  Multichromatic Floquet driving has also been applied recently to manipulate topological quantum states~\cite{MultichromaticFloquet1,MultichromaticFloquet2}. In the following, we discuss how an effective quantum dissipation can emerge in a time coarse-grained (TCG) dynamics. For this purpose, we exploit a timescale separation formalism \cite{Rahav03,Goldman14} for a multichromatic driving, and infer an effective Master equation for the TCG matrix density with well-controlled approximations, and valid over a long time interval.

Consider a quantum system driven by a time-independent Hamiltonian $\hat{H}_0$ and a Floquet Hamiltonian $\hat{H}_{F}(t) = \sum_m \hat{V}_m e^{i \omega_m t}+ h.c.$. The corresponding evolution operator can be recast as the product of three unitary transforms involving separately either slow or fast-evolving operators~\cite{Rahav03,Goldman14}:

\begin{equation}
	\label{eq:generalframeworkevolution}
	\hat{U}(t,t_0) = e^{- i \hat{K}(t)}  \hat{U}^{\rm eff}(t) e^{ i \hat{K}(t_0)}, 
\end{equation}
where $\hat{U}^{\rm eff}(t)=\mathcal{T}\left[ e^{- i \int_{t_0}^t dt'\hat{H}^{\rm eff}(t') } \right]$ accounts for the slow dynamics under the effective Hamiltonian $\hat{H}^{\rm eff}(t)$ ($\mathcal{T}$ is the time ordering operator), while the terms involving the kick operator, $\hat{K}(t)$, contain the fast sinusoidal time-dependence. The Floquet frequencies $\omega_m$ are assumed to be much larger than the eigenfrequencies of $\hat{H}_0$ and $ \hat{V}_m$:  $\varepsilon= \Omega/\omega \ll 1$ with $\Omega= \max_m \{|| \hat{H}_0 ||,|| \hat{V}_m || \}$ and $\omega = \min_m \{ \omega_m \}$. This frequency hierarchy is used to expand $\hat{H}^{\rm eff}(t)= \sum_{n=0}^{+\infty} \hat{H}^{\rm eff}_n(t)$ and $\hat{K}(t)= \sum_{n=1}^{+\infty} \hat{K}_n(t)$ where $||\hat{K}_n(t) || = O \left( \frac {\Omega^n} {\omega^n} \right)$ and $||\hat{H}^{\rm eff}_n(t)||=O \left( \frac {\Omega^{n+1}} {\omega^n} \right).$ 

The instantaneous quantum state $| \psi(t) \rangle$ undergoes a unitary evolution with a fast time-dependence. However, the evolution of the TCG density matrix $\overline{\rho}(t)= \overline{| \psi(t) \rangle \langle \psi(t) |}$ is in general non-unitary. The considered TCG procedure works as a low-pass filter in frequency space involving a cutoff frequency $\omega_c$: $\overline{\hat{O}}(t)= \frac {1} {\sqrt{2 \pi}} \int_{- \omega_c}^{\omega_c} \hat{O}(\omega) e^{-i \omega t} d \omega ,$ where $\hat{O}(\omega)= \frac {1} {\sqrt{2 \pi}} \int_{- \infty}^{+ \infty} \hat{O}(t) e^{i \omega t} dt$ is the Fourier transform of the considered operator $\overline{\hat{O}}(t)$.  The cutoff frequency $\omega_c$ is chosen to leave invariant the slow  Hamiltonian dynamics, i.e. $\overline{e^{\pm i \hat{H}_0 t}}=e^{\pm i \hat{H}_0 t}$, while filtering out the fast Floquet frequencies $\forall m \: \overline{e^{\pm i \omega_m t}}=0$ ($\overline{\hat{H}_F(t)}=0$). Finally, we assume that for the slow operators considered  below (such that $\overline{\hat{O}_{\rm slow}(t)}=\hat{O}_{\rm slow}(t)$), one always has $\overline{\hat{O}_{\rm slow}(t)\hat{O}(t)}=\hat{O}_{\rm slow}(t) \overline{\hat{O}(t)}$ and  $\overline{\hat{O}(t) \hat{O}_{\rm slow}(t)}=\overline{\hat{O}(t)} \hat{O}_{\rm slow}(t)$. This property is fulfilled if the $\hat{O}_{\rm slow}$ operator oscillates at frequencies $\omega_{\rm slow} \ll  \omega_c$ and if the $\hat{O}(t)$ operator does not have frequencies nearby the cutoff $\omega_c$. These assumptions are realistic for a sufficient large separation between the slow and fast timescales.

We now proceed to derive an effective Master equation for the TCG density matrix. From Eq.~\eqref{eq:generalframeworkevolution}, we obtain
 $\overline{\rho}(t)=\overline{e^{- i \hat{K}(t)} \rho_e(t) e^{i \hat{K}(t)}}$ with  $\rho_e(t)= \hat{U}^{\rm eff}(t) e^{ i \hat{K}(t_0)} | \psi (t_0) \rangle \langle \psi(t_0)| e^{ - i \hat{K}(t_0)} \hat{U}^{\rm eff}(t)^{\dagger}$ evolving under the effective Hamiltonian $\hat{H}^{\rm eff}(t)$. By construction of the effective Hamiltonian~\cite{Rahav03,Goldman14}, the density matrix $\rho_e(t)$ follows slow dynamics and fulfills $\overline{\rho_e(t)}=\rho_e(t)$. We subsequently expand the fast unitary transforms $e^{\pm i \hat{K}(t)}$ in terms of the small parameter $\varepsilon=\Omega/\omega$. The TCG density matrix then reads
 \begin{equation}
 	\label{eq:expansion0}
 \overline{\rho}(t)=\rho_e(t)+\sum_{n=1}^N \overline{\delta \rho^{(m)}}(t) +O \left(  \varepsilon^{N+1} \right)
 \end{equation}
Each term $\overline{\delta \rho^{(m)}(t)}$ represents a correction of order $O(\varepsilon^m)$ and depends linearly on the density matrix $\rho_e(t)$. In order to derive these corrections, one needs explicit expressions for the fast kick operators $\hat{K}_m(t)$. These are used to cancel the fast time-dependence in the effective Hamiltonian, and can be obtained at each order through a systematic procedure \cite{Rahav03,Supplementary}. For instance, $\hat{K}_1(t)$ fulfills $\dot{\hat{K}}_1(t) = \hat{H}_F(t)$ and reads $\hat{K}_1(t)= \sum_m \frac {1} {i \omega_m} (\hat{V}_m e^{i \omega_m t}- h.c)$~\cite{Supplementary}. The lowest-order correction
is of second-order as $\delta \overline{\rho^{(1)}}(t)= - i [ \overline{ \hat{K}_1(t)} , \rho_e(t)] =0$ and is given by $\delta \overline{\rho^{(2)}}(t)= - \frac 1 2 \{ \overline{\hat{K}_1(t)^2} , \rho_e(t) \}	+  \overline{\hat{K}_1(t) \rho_e(t) \hat{K}_1(t)}$. 
An effective equation for the time-averaged density matrix is obtained by taking the time derivative of Eq.~(\ref{eq:expansion0}). Special care is, however, needed in order to gather consistently corrections to the same order. For instance, the contribution $\overline{\delta \rho^{(2)}(t)}$ involves a product of fast-evolving ($\hat{K}_1(t)$) and slow-evolving (the density matrix $\rho_e(t)$) functions. When applied to the latter, the time derivative yields terms which are smaller by one order in the small parameter $\varepsilon$. This leads us to distinguish the slow and fast time dependence by setting $\tau$ and $t$ for the corresponding time variables, with $\partial_\tau=O(\Omega)$ and $\partial_t=O(\omega)$, similarly to the two-timing technique~\cite{Footnote1,StrogatzBook}. We note $\overline{\delta \rho^{(m)}}(t,\tau)$ the corresponding corrections to the density matrix, so that $\overline{\delta \rho^{(2)}}(t,\tau)= - \frac 1 2 \{ \overline{\hat{K}_1(t)^2} , \rho_e(\tau) \}	+  \overline{\hat{K}_1(t) \rho_e(\tau) \hat{K}_1(t)}$.  Furthermore, we assume that the Floquet frequencies $\omega_m$ are grouped in a narrow bandwidth, i.e. $\forall (m,n) \: |\omega_m-\omega_n| < \omega_c \ll \omega.$ The third-order correction $\overline{\delta \rho^{(3)}}(t)$ involves only contributions from the two lowest-order fast operators $\{ \hat{K}_1(t), \hat{K}_2(t) \}$ as the time-averaging eliminates the isolated contribution of the fast operator $\hat{K}_3(t)$. Cubic terms $\hat{K}_1(t)^3 \rho_e(t), \hat{K}_1(t)^2\rho_e(t)\hat{K}_1(t),... $ do not contain low-frequency harmonics and thus disappear upon time-averaging. One obtains
$\overline{\delta \rho_3}(t,\tau) = \overline{\hat{K}_1(t) \rho_e(\tau) \hat{K}_2(t)}+ \overline{\hat{K}_2(t) \rho_e(\tau) \hat{K}_1(t)} - \frac 1 2 \{  \overline{\{\hat{K}_1(t), \hat{K}_2(t) \}} , \rho_e(\tau) \}$. The complete effective Master equation can be written to second order as  $\frac {\partial} {\partial t} \overline{\rho} = - i [\hat{H}^{\rm eff}, \rho_e] + \partial_t \overline{\delta \rho^{(2)}(t,\tau)} + \partial_\tau \overline{\delta \rho^{(2)}(t,\tau)}+ \partial_t \overline{\delta \rho^{(3)}(t,\tau)}+O(\Omega \varepsilon^3 )$. At the second order expansion, $\rho_e(t)=\overline{\rho}(t)-\overline{\delta \rho^{(2)}}(t)+O(\varepsilon^3)$ in the unitary term of the r.h.s., but $\rho_e(t) \simeq \overline{\rho}(t)$  is sufficient in the Lindblad terms.

We eventually obtain  the effective Master equation for the TCG density matrix which constitute the central result of this article \cite{Supplementary}:
\begin{equation}
	\label{eq:FullMasterEquation}
\frac {\partial \overline{\rho} } {\partial t}  =   -i [ \hat{H}_{\rm eff}, \overline{\rho} ] +\mathcal{L}_2^{FF}[\overline{\rho}] +\mathcal{L}_2^{FSF}[\overline{\rho}]+O(\Omega \varepsilon^3)
\end{equation}
with
\begin{eqnarray}
	\mathcal{L}_2^{FF}[\overline{\rho}] \! \! \! & \! = \! &	\! \! \!
		\sum_{m<n} \! \frac {4 \: \overline{\sin (\Delta \omega_{nm}t)}} {\omega_{mn-}} \! \mathcal{D}[\hat{V}_m^{\dagger},\hat{V}_n][\overline{\rho}], \label{l2c} \\
\mathcal{L}_2^{FSF}[\overline{\rho}]  \! \! & \! \! = \! \! & \! \!  i  \sum_{m,n} \left[   \frac {1} {\omega_n^2} \mathcal{D}[\hat{V}_m,[\hat{V}_n^{\dagger},\hat{H}_0]][\overline{\rho}] \right. \nonumber \\
& + & \left. \frac {1} {\omega_m^2} \mathcal{D}[\hat{V}_n^{\dagger},[\hat{V}_m,\hat{H}_0]][\overline{\rho}] \right] e^{i (\omega_m-\omega_n)t},  \label{l2nc}
\end{eqnarray}
with $\Delta \omega_{nm}= \omega_n-\omega_m$, $1/\omega_{mn-}=\frac 1 2 (1/\omega_{m}-1/\omega_{n})$ and $\mathcal{D}[\hat{V},\hat{V}'][\overline{\rho}]  =    \frac 1 2  \{ \{ \hat{V}, \hat{V}' \} , \overline{\rho} \} -\hat{V} \overline{\rho} \hat{V}' \!- \! \hat{V}' \overline{\rho} \hat{V} $.

This effective Master equation contains two non-unitary contributions encapsulated in the Linblad-like terms $\mathcal{L}_2^{FF}[\overline{\rho}]$ and $\mathcal{L}_2^{FSF}[\overline{\rho}]$  provided that the Floquet Hamiltonian contains at least two different frequencies $\{ \omega_m , \omega_n \}$ close enough to ensure  $\overline{e^{\pm i (\omega_m-\omega_n)t}}=e^{\pm i (\omega_m-\omega_n)t}$.  Under this assumption, the beat notes between these Floquet modes generate tunable non-Hermitian contributions to the time-averaged dynamics when the inequalities $|\omega_m-\omega_n| < \omega_c \ll \omega$ are satisfied. The non-unitary operator $\mathcal{L}_2^{FF}[\overline{\rho}]$ is bilinear in the Floquet operators and scales as $1/\omega_{mn-} \simeq |\Delta \omega_{mn}|/\omega^2$. For usual situations where  $|\Delta \omega_{mn}| \leq \Omega,$ which corresponds to dissipation terms oscillating at a comparable pace (or slower) as the effective Hamiltonian dynamics, the non-unitary operators of Eq.~\eqref{eq:FullMasterEquation} are of second-order. The extra contribution~(\ref{l2nc}) arises when $[\hat V_m, \hat{H}_0]\neq 0$, and akes into account the interaction between slow and fast quantum dynamics in the resulting time-averaged evolution. 

The effective Master equation derived in the present framework is valid over an arbitrary long time interval. This is an essential benefit from our approach based on the exact expression~\eqref{eq:generalframeworkevolution} followed by an expansion in terms of the Floquet frequencies. We obtain (see \cite{Rahav03,Goldman14} and the SM~\cite{Supplementary}) $\hat{H}^{\rm eff}_0=\hat{H}_0,$ $\hat{H}^{\rm eff}_1 = \frac 1 2 \sum_{m,n} \left(\frac {1} {\omega_m}  + \frac {1} {\omega_n} \right) [\hat{V}_m, \hat{V}_n^{\dagger}] e^{i (\omega_m-\omega_n)t}$ and $\hat{H}^{\rm eff}_2 = \frac 1 2 \sum_{m,n} \left( \frac {1} {\omega_m^2} [[\hat{V}_m, \hat{H}_0], \hat{V}_n^{\dagger}]+ \frac {1} {\omega_n^2} [ [\hat{V}_n^{\dagger}, \hat{H}_0], \hat{V}_m] \right) e^{i (\omega_m-\omega_n)t}$.
At the considered second-order and for Floquet frequencies taken in a narrow bandwidth, kick operators must be grouped pairwise in order to generate low-frequency harmonics that survive the time-averaging. This is why the bichromatic case considered below, contains the phenomenology of the non-unitary effects that arise in any multichromatic Floquet driving. 

\begin{figure}
	\centering
	\includegraphics[width=8.5 cm]{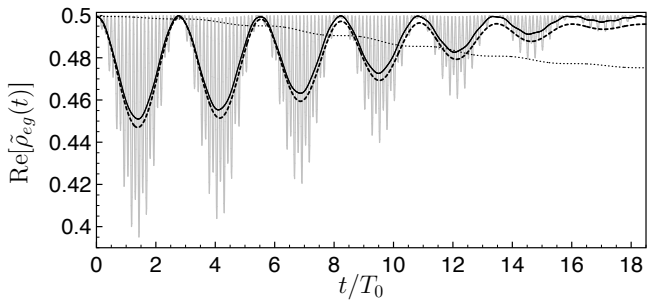}  
	\caption{Quantum dynamics in the interaction picture with $\hat{H}_0 \propto \sigma_z$ and $\hat{V}_m \propto \sigma_x$: Instantaneous density matrix profile ${\rm Re}[\tilde{\rho}_{eg}(t)]$ as a function of time (solid gray line)  with $\tilde{\rho}(t)=U^{\rm eff}(t)^{\dagger} \rho(t) U^{\rm eff}(t)$. Coarse-grained density matrix coherence  ${\rm Re}[\overline{\tilde{\rho}}_{eg}(t)]$ as a function of time obtained from a time-averaging of the instantaneous solution $\tilde{\rho}(t)$ (solid black line), or from a resolution of the full effective Master equation (dashed black line), of an effective Master equation without the contribution  $\mathcal{L}^{FSF}[\overline{\tilde \rho}]$ (see text) to the quantum dissipation term (dotted line). Results obtained for the constant/Floquet Hamiltonians $\hat{H}_0=\omega_0 \sigma_z$, $\hat{H}_F(t)= \Omega_1 \sigma_x e^{ i \omega_1 t} +\Omega_2  \sigma_x e^{i \omega_2 t}+h.c.$, and with an initial density matrix $\rho_0=| \psi_+ \rangle \langle \psi_+|$ where $| \psi_+ \rangle = \frac {1} {\sqrt{2}} (| e \rangle +| g \rangle )$. Parameters used: $\omega_0=0.1 \times (2 \pi)/ T_0$ and $\Omega_1=\Omega_2=2/T_0$. Floquet frequencies: $\omega_1=4 \times (2 \pi) /T_0 $, $\omega_2=\omega_1+ \Delta \omega_{21} $ with  $\Delta \omega_{21}=0.025 \times (2 \pi)  /T_0$, and small parameter $\epsilon =0.1$. TCG realized with the cut-off frequency $\omega_c = 2 \times (2\pi)/T_0$.}	
	\label{fig:NonCommutative}
\end{figure}

As a first example, we consider a two-level system with  $\hat{H}_0=\omega_0 \sigma_z$, and the Floquet operators $\hat{V}_m= \Omega_m \sigma_x$ for $m=1,2$ ($\sigma_{x,y,z}$ are the Pauli matrices, and we set here $\Omega_{1,2}=\Omega>0$). This choice yields $\mathcal{L}_2^{FF}[\overline{\rho}]= 8 \Omega^2  \left(  \sin (\Delta \omega_{21} t) / \omega_{12-} \right) (\overline{\rho}- \sigma_x \overline{\rho} \sigma_x )$ and $\mathcal{L}^{FSF}[\overline{\rho}]= - 8 \omega_0 \Omega^2 \left( \frac {1} {\omega_1^2}+\frac {1} {\omega_2^2} \right) \cos^2 ( \frac 1 2 \Delta \omega_{21} t)  ( \sigma_x \overline{\rho} \sigma_y+ \sigma_y \overline{\rho} \sigma_x )$. The effective Hamiltonian contributions are $\hat{H}^{\rm eff}_1=0$ and $\hat{H}^{\rm eff}_2= - 8 \omega_0 \Omega^2 \left( \frac {1} {\omega_1^2}+\frac {1} {\omega_2^2} \right) \cos^2 ( \frac 1 2 \Delta \omega_{21} t)  \sigma_z$. To emphasize the role of the dissipative dynamics, we provide hereafter the quantum evolution within the interaction picture with respect to the second-order effective Hamiltonian $\hat{H}^{\rm eff}=\hat{H}_0+\hat{H}^{\rm eff}_2$. Figure \ref{fig:NonCommutative} pictures the time evolution of the instantaneous density matrix coherence $\tilde{\rho}_{eg}(t)$ with $\tilde{\rho}(t)= \hat{U}^{\rm eff}(t)^{\dagger} \rho(t) \hat{U}^{\rm eff}(t)$. We express all time-related quantities using an arbitrary time unit $T_0$. We also provide the TCG evolution using a convolution with the sinus cardinal function $f(t)=\sin (\omega_c t) / (\pi t).$ We subsequently compare this time-averaged density matrix $\overline{\tilde \rho}(t)= \int dt' f(t'-t) \tilde \rho(t)$ with the predictions of the effective Master equation with the initial condition $\overline{\tilde \rho}_0= e^{i \hat{K}_1(t_0)} \left( \int_{- \infty}^{+\infty} dt f(-t) \tilde \rho(t) \right) e^{-i \hat{K}_1(t_0)}$.
 We have also added the prediction from the Master equation in the absence of the  $\mathcal{L}^{FSF}[\overline{\tilde \rho}]$ term, i.e. as derived in \cite{Gamel10}. This latter approach, based on a Dyson perturbative expansion, yields an effective Master equation whose validity is restricted by construction to a very short time interval, and to moderate dissipation strengths. As a matter of fact, these assumptions can be overly restrictive, as a long duration is needed for moderate dissipation to alter significantly the dynamics of a given quantum system. 

Our second example illustrates the emulation of phase noise in the time-averaged quantum dynamics. In NMR, such a dissipative physical mechanism results from fluctuations of the magnetic field. The phenomenological equation for  the average spin dynamics, $\dot{\mathbf{M}} =  \gamma \mathbf{M} \times \mathbf{B} + (M_0- M_z)/T_1 \hat{\mathbf{z}} - \mathbf{M}_{\perp}/T_2$, accounts for the dissipative effects through two times $T_1$ and $T_2$, associated respectively to the longitudinal ($M_z$) and transverse ($\mathbf{M}_{\perp}$) relaxations. In terms of the density matrix, the former corresponds to the population difference $\rho_{ee}-\rho_{gg}$ while the latter involves the density matrix coherences $\rho_{eg},\rho_{ge}$. The phase noise is accounted for with decay times $T_1=+\infty$ and $T_2=1/\gamma$~\cite{Levitt08}. The Master equation that models the phase noise reads $\partial_t \rho = - i [\hat{H}_0,\rho] +\frac {\gamma} {2} \mathcal{L}_{\rm phase}[\rho]$ with the Liouvillian $\mathcal{L}_{\rm phase}[\rho]=\sigma_z \rho \sigma_z - \frac 1 2 \{ \sigma_z \sigma_z, \rho\} $.  

To emulate such a dissipative dynamics, we consider a bichromatic driving with $\hat{H}_0= \omega_0 \sigma_z$ and $\hat{V}_m= \Omega_m \sigma_z$ for $m=1,2$. In this particular case, the contribution of the $\mathcal{L}^{FSF}[\overline{\rho}]$ term vanishes and the resulting Master equation coincides with the desired form with $\gamma(t)= -16 {\rm Re}[\Omega_1^* \Omega_2] \sin (\omega_2-\omega_1)t/\omega_{12-}$ (the contribution of the $ \mathcal{D}[\hat{V}_m^{\dagger},\hat{V}_n][\rho]$ term is proportional to a Linbladian operator $\mathcal{L}_{\rm phase}[\rho]$). Here, the coefficient $\gamma$ has a time-dependent value and alternates between regimes of gain  ($\gamma <0$) and damping ($\gamma >0$). Setting very close and non-commensurate frequencies $\omega_1$ and $\omega_2$ enables to accumulate decoherence (or gain) over a significant time interval. 

\begin{figure}
	\centering
	\includegraphics[width=8.5 cm]{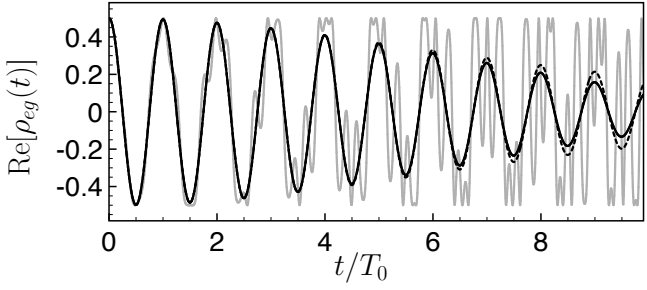}  
	\caption{Emulation of Phase Noise: Results of the effective Master equation vs full quantum evolution: Instantaneous density matrix coherence ${\rm Re}[\rho_{eg}(t)]$ as a function of time (solid gray line), time coarse-grained coherence ${\rm Re}[\overline{\rho}_{eg}(t)]$ (solid black line) and the density matrix coherence ${\rm Re}[\overline{\rho}_{eg}(t)]$  (dashed black line) obtained from the effective Master equation (Eq.~\ref{eq:FullMasterEquation}). Parameters used: constant 
		and Floquet Hamiltonians corresponding to $\hat{H}_0=\omega_0 \sigma_z$ and $\hat{H}_F(t)= \Omega_1 \sigma_z e^{ i \omega_1 t} +\Omega_2  \sigma_z e^{i \omega_2 t}+h.c.$ with  $\omega_0=0.5 \times (2 \pi)/ T_0$ and Floquet terms in opposite phase $\Omega_1=-\Omega_2=7/T_0$. We have used noncomensurate Floquet frequencies $\omega_1=\sqrt{10} \times (2 \pi) /T_0 $, $\omega_2=\omega_1+ \Delta \omega_{21} $ with  $\Delta \omega_{21}=0.025 \times (2 \pi)  /T_0$, yielding the small parameter $\varepsilon=0.35$. The initial density matrix ($\rho_0$) and cut-off ($\omega_c$) frequencies are identical to Fig.1.}
	\label{fig:PhaseNoise}
\end{figure}

As previously, we validate numerically our findings by resolving the full unitary quantum dynamics driven by the Hamiltonian $\hat{H}(t)=\hat{H}_0+\hat{H}_F(t)$. In Fig.~\ref{fig:PhaseNoise}, we illustrate our results. The seemingly erratic oscillations of the instantaneous density matrix coherence depicted in Fig.~\ref{fig:PhaseNoise} \cite{Supplementary} generate a TCG dynamics that follows very accurately the effective Master equation, i.e. the one of a damped Rabi oscillation. 
This averaging effect on the Floquet-induced peaks is reminiscent of the averaging on individual stochastic trajectories involving  quantum jumps in the Monte Carlo wave function formalism \cite{PRLJeanYvan}.
Floquet-induced peaks accumulate periodically at a pace determined by the beat frequency $\Delta \omega_{21}$ between the two involved Floquet modes. This periodic increase/decrease of sharp peaks provokes an oscillation of the effective damping rate $\gamma(t)$ at the same frequency $\Delta \omega_{21}$. An initial loss (gain) phase can be obtained by setting a specific phase difference $\phi$ between the two Floquet modes. By convention we use $\Omega_1 \in \mathbf{R}^+$ and $\Omega_2=|\Omega_2| e^{i \phi} $, with the Floquet frequencies ordered with their labels $\omega_n > \omega_m$ if $n>m$. The choice $\Omega_2=-\Omega_1$ gives the decoherence pictured in Fig.~\ref{fig:PhaseNoise}. 


Our framework provides a very accurate approximation of the full time-averaged dynamics in this second example. This is not obvious, as Eq.~\eqref{eq:FullMasterEquation} is a mere second-order approximation, and discards several contributions associated to the higher-order kick operators $\hat{K}_m(t)$. Actually,
 the operators $\hat{K}_m(t)$ vanish here for $m \geq 2$ as a result of the commutation between the Floquet and time-independent Hamiltonians. Thus, the expansion of the unitary operators $e^{\pm i \hat{K}(t)}$ boils down to a simple power expansion in the operator $\hat{K}_1(t)$. Furthermore, odd powers of $\hat{K}_1(t)$ do not generate any low-frequency harmonics, and the effective Master equation only receive contributions from even-order terms. Incidentally, the 4th-order contribution also cancels~\cite{Supplementary}. In the special case of commuting operators, Eq.~\eqref{eq:FullMasterEquation} is thus accurate up to the 5th-order, which explains the remarkable agreement between the approximate effective Master equation~\eqref{eq:FullMasterEquation} and the full quantum dynamics, which still holds for moderate values of the parameter $\varepsilon$ ($\varepsilon$=0.35 in Fig.~\ref{fig:PhaseNoise}).

In our third example, we propose to emulate a quantum dynamics reminiscent of incoherent emission/absorption processes in the TCG evolution of a two-level system. These processes are described respectively by the Liouvillians $\mathcal{L}_{\rm em}[\rho]= \sigma_- \rho \sigma_+ - \frac 1 2 \{ \sigma_- \sigma_+, \rho \}$ and $\mathcal{L}_{\rm ab}[\rho]= \sigma_+ \rho \sigma_- - \frac 1 2 \{ \sigma_+ \sigma_-, \rho \}$, where $\sigma_+=|e\rangle \langle g|$ and  $\sigma_-=\sigma_+^{\dagger}$. By symmetry of the dissipative term $\mathcal{D}[\hat{V},\hat{V}'][\rho]$ in the effective Master equation, if the TCG dynamics contains the Liouvillian $\mathcal{L}_{\rm em}[\rho]$, it also contains the Liouvillian $\mathcal{L}_{\rm ab}[\rho]$ associated to the reverse process. This regime illustrates, for example, the dynamics of a two-level atom illuminated by an intense light field, so that stimulated emission predominates over spontaneous emission \cite{LivreCCTDGO}. In this case, the emission/absorption rates are approximately equal $\gamma_{\rm em} \simeq \gamma_{\rm ab} \simeq \gamma$.

\begin{figure}
	\centering
	\includegraphics[width=8.5 cm]{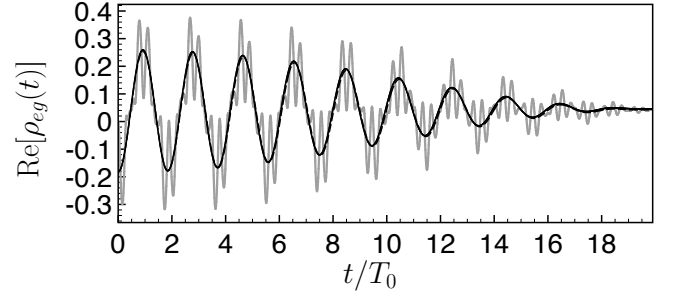}  
	\caption{Incoherent absorption/emission in the time-averaged dynamics: Instantaneous density matrix coherence ${\rm Re}[\rho_{eg}(t)]$ as a function of time (solid gray line),  time-averaged coherence ${\rm Re}[\overline{\rho}_{eg}(t)]$ (solid black line) vs density matrix coherence ${\rm Re}[\overline{\rho}_{eg}(t)]$  (dashed black line) obtained from the effective Master equation. Parameters: initial density matrix $\rho_0=| e \rangle \langle e|$, constant 
		and Floquet Hamiltonians $\hat{H}_0=\omega_0 \sigma_x$, $\hat{H}_F(t)= \Omega_1 \sigma_+ e^{ i \omega_1 t} +\Omega_2  \sigma_+ e^{i \omega_2 t}+h.c.$ with the frequency $\omega_0=0.25 \times (2 \pi)/ T_0$ and in-phase Floquet terms s.t. $\Omega_{1,2}=2/T_0$. Small parameter $\epsilon =0.1$. Floquet ($\omega_{1,2}$) and cutoff ($\omega_c$) frequencies are identical to those of Fig.~2.}		
	\label{fig:emissionabsorption}
\end{figure}

To produce an analog of this dissipation, we take the time-independent $\hat{H}_0= \omega_0 \sigma_x$ and Floquet Hamiltonians with $\hat{V}_m= \Omega_m \sigma_+$ for $m=1,2$ ($\Omega_{1,2}=\Omega>0$). With this choice, the bilinear term $\mathcal{L}_2^{FF}[\overline{\rho}]$ accounts for these two incoherent processes as  $\mathcal{L}_2^{FF}[\overline{\rho}]= \gamma(t)  (\mathcal{L}_{\rm em}[\overline{\rho}]+\mathcal{L}_{\rm ab}[\overline{\rho}])$ with the time-dependent effective emission/absorption rate $\gamma(t)= - 4  \Omega^2 \sin ( \Delta \omega_{21} t) / \omega_{12-}.$  The remaining contribution reads $\mathcal{L}_2^{FSF}[\overline{\rho}]= -  \omega_0 \Omega^2 \left(\frac {1} {\omega_1^2} + \frac {1} {\omega_2^2}  \right) \cos^2 ( \frac 1 2 \Delta \omega_{21} t) (\sigma_y \rho \sigma_z+ \sigma_z \rho \sigma_y) + \Omega O( \varepsilon^3)$. One finds the effective Hamiltonian corrections
$\hat{H}^{\rm eff}_1 = 2 \Omega^2 \left(\frac {1} {\omega_1} + \frac {1} {\omega_2} \right) \cos^2 ( \frac 1 2 \Delta \omega_{21} t) \sigma_z$ and $\hat{H}^{\rm eff}_2 = - 2 \omega_0  \Omega^2 \left(\frac {1} {\omega_1^2} + \frac {1} {\omega_2^2}  \right) \cos^2 ( \frac 1 2 \Delta \omega_{21} t) \sigma_x + \Omega O( \varepsilon^3)$. In Fig.~\ref{fig:emissionabsorption}, we plot as a function of time the exact instantaneous density matrix coherence and its corresponding TCG evolution. We observe an excellent agreement with the prediction of the effective Master equation (the two curves are almost perfectly superposed).

More generally, our approach enables one to emulate a Lindblad Master equation of the form $\dot{\rho}= - i [\hat{H},\hat{\rho}]+\sum_{m=1}^N \gamma_m \left[   \hat{L}_m \rho \hat{L}_m^{\dagger}+\hat{L}_m^{\dagger} \rho \hat{L}_m-\frac 1 2 \{ \{ \hat{L}_m , \hat{L}_m^{\dagger} \} ,\rho \} \right]$, i.e. involving, for each quantum jump operator $\hat{L}_m$, the reverse jump $ \hat{L}_m^{\dagger}$ at the same rate $\gamma_m$~\cite{Footnote2}. It is approximately generated by the Floquet Hamiltonian $\hat{H}_F(t)=\sum_{m=1}^N \Omega_m \hat{L}_m \left( e^{i \omega_m t} +e^{i (\omega_m+ \Delta \omega_m )t+i\varphi_m} \right) + h.c.$ with well-separated pairs of close frequencies $\{ \omega_m,\omega_m+\Delta \omega_m \}$, such that $\Delta \omega_m \ll \omega_c$ and  $|\omega_m-\omega_n| > \omega_c$ for $m \neq n$ in order to avoid crossed terms involving different pairs of jump operators.  With these assumptions, the operator $\mathcal{L}_2^{FF}[\overline{\rho}]$~\eqref{l2c} takes the desired form. Interestingly,  the effective time-dependent rates $\gamma_m(t) \simeq -4 \left( |\Omega_m |^2  \Delta \omega_m/  \omega_m^2 \right) \sin (\Delta \omega_m t+ \varphi_m)$ can be shaped independently by a suitable choice of the Rabi pulsations ($\Omega_m$), frequency ($\Delta \omega_m$) and phase ($\varphi_m$) differences. Regarding the $\mathcal{L}_2^{FSF}[\overline{\rho}]$ term, its contribution can be attenuated by an appropriate choice of $\hat{H}_0$, e.g. the third example detailed above.

In summary, we have used the formalism of kick operators and effective Hamiltonians to derive an effective Master equation for the TCG dynamics in a multichromatic Floquet system. Our treatment, based on a perturbative expansion in terms of powers of kick operators, holds in the long-time limit. The beat modes between pairs of Floquet frequencies generate effective quantum dissipation that results from a blurring of the fast instantaneous motion.  Different Floquet Hamiltonians and time-averaging procedures can be considered to emulate a wide range of dynamics involving gains or losses. Our approach paves the way for quantum simulations based on Floquet-engineered non-unitary dynamics.

\textit{Acknowledgments}. The authors thank Jean Dalibard for useful comments.  F.I. was supported by the Brazilian agencies CNPq (310265/2020-7), CAPES and FAPERJ (210.296/2019). This work was supported by the CAPES-PRINT Program and by INCT-IQ (465469/2014-0).

\newpage

\begin{widetext}

\section*{SUPPLEMENTAL MATERIAL}


In Sec.~\ref{sec1}, we provide the derivation of the kick operators and effective Hamiltonians to the second order. In  Sec.~\ref{sec2}, we establish the effective Master Equation that governs the time-coarsed grained density matrix to the second order. In Sec.~\ref{sec3}, we exemplify the calculations of the higher order corrections to the effective Master equation for the applications detailed in the main text.

\section{THE KICK OPERATORS AND EFFECTIVE HAMILTONIANS TO second ORDER}
\label{sec1}


The exact Hamiltonian under consideration reads $\hat{H}(t)=\hat{H}_0+\hat{H}_F(t)$ with $\hat{H}_F(t)= \sum_m \hat{V}_m e^{i \omega_m t } + h.c.$. We assume that all Floquet frequencies $\omega_m$ belong to a narrow bandwidth, i.e. fulfill $\forall m \: ,\forall n, \: |\omega_m -\omega_n|  \ll \omega_c \ll \omega$. Following the procedure of Refs.\cite{Rahav03,Goldman14}, we search for a unitary operator $e^{ i \hat{K}(t)}$ such that the state expressed in the new gauge $| \phi(t) \rangle = e^{i K(t)} | \psi(t) \rangle $ follows a slow dynamics. The Hamiltonian in the new gauge frame is given by
\begin{equation}
	\label{eq:transformedHamiltonian}
	\hat{H}^{\rm eff}(t) = e^{i \hat{K}(t)} \hat{H}(t) e^{-i \hat{K}(t)}+ i \frac {\partial e^{i \hat{K}(t)}} {\partial t}  e^{-i \hat{K}(t)},
\end{equation}
and must be such that $\overline{H_{\rm eff}(t)}=H^{\rm eff}(t)$ at any time $t$. For the considered Floquet frequencies, one has $\overline{e^{i \pm (\omega_m-\omega_n)t}}=e^{i \pm (\omega_m-\omega_n)t}$ and $\overline{e^{ \pm i \omega_m t}}=\overline{e^{ \pm i (\omega_m+\omega_n) t}}=0.$ Thus, only terms rotating at a 
difference between two Floquet frequencies (or constant terms) will contribute to the effective Hamiltonian $H^{\rm eff}(t)$.   In this section, we determine iteratively the first contributions to the expansion $\hat{H}^{\rm eff}(t)= \sum_{n=0}^{+\infty} \hat{H}^{\rm eff}_n(t)$ and $\hat{K}(t)= \sum_{n=1}^{+\infty} \hat{K}_n(t)$ using the identities provided by the  from the Baker-Campbell-Hausdorff formula:
\begin{eqnarray}
	e^{i \hat{K}(t)} \hat{H}(t) e^{-i \hat{K}(t)} & = & \hat{H}(t) + i [\hat{K}(t), \hat{H}(t)] - \frac 1 2 [ \hat{K}(t), [\hat{K}(t), \hat{H}(t)] ] \nonumber \\ &- &  \frac {i} {6} [ \hat{K}(t), [ \hat{K}(t), [ \hat{K}(t), \hat{H}(t)]]] +... \label{eq:Id1} \\
	\left( \frac {\partial e^{i \hat{K}(t)}} {\partial t} \right)  e^{ - i \hat{K}(t)} & = & i \frac {\partial \hat{K}} {\partial t} - \frac 1 2 \left[ \hat{K}(t), \frac {\partial K} {\partial t} \right] - \frac i 6 \left[ \hat{K}(t), \left[ \hat{K}(t), \frac {\partial K} {\partial t} \right] \right] +... \label{eq:Id2}
\end{eqnarray}
In the following, we explicitly use the fact that $\hat{H}^{\rm eff}_n(t), \hat{K}_n(t)$ and $\frac {\partial} {\partial t} \hat{K}_{n+1}$ are of the same order $O(\varepsilon^n)$ where $\varepsilon=\Omega/\omega$. The zero-order contribution is obtained by taking $ e^{i \hat{K}(t)} \hat{H}(t) e^{-i \hat{K}(t)} = H(t)+O(\varepsilon)$. Using Eqs.~(\ref{eq:transformedHamiltonian}) and (\ref{eq:Id2}), we find 
\begin{equation}
	\hat{H}^{\rm eff}_0(t) = \hat{H}_0 +\hat{H}_F  - \frac {\partial \hat{K}_1} {\partial t}. \nonumber
\end{equation}
From now on, we remove the explicit time-dependence of the operators on the r.h.s. to alleviate notations when needed. As $\overline{\hat{H}_0}=\hat{H}_0$ and $\overline{\hat{H}_F(t)}=0$, we set $	\hat{H}^{\rm eff}_0(t)=\hat{H}_0 $ and
\begin{equation}
	\label{eq:K1derivative}
	\frac {\partial \hat{K}_1} {\partial t}=\hat{H}_F.
\end{equation}
The kick operator $\hat{K}_1(t)$ removes all the fast time-dependence from the effective Hamiltonian, and can be chosen as
\begin{equation}
	\label{eq:K1}
	\hat{K}_1(t) =  \sum_m \frac {1} {i \omega_m} \left( \hat{V}_m e^{ i \omega_m t } - \hat{V}_m^{\dagger} e^{- i \omega_m t }  \right),
\end{equation}
up to an arbitrary constant operator.

To obtain the result to next-order, we introduce the two lowest-order kick operators $\hat{K}_{1,2}(t)$ into Eqs.~(\ref{eq:transformedHamiltonian},\ref{eq:Id1},\ref{eq:Id2}). We find
\begin{equation}
	H^{\rm eff}_1(t)= i 	[ \hat{K}_1, \hat{H} ]   - \frac i 2 \left[ \hat{K}_1, \frac {\partial \hat{K}_1} {\partial t} \right] - \frac {\partial \hat{K}_2} {\partial t}, \nonumber
\end{equation}
which can be recast thanks to Eq.~\eqref{eq:K1derivative} as
\begin{equation}
	\label{eq:Heff1def}
	\hat{H}^{\rm eff}_1(t)=  i 	[ \hat{K}_1, \hat{H}_0 ]+ \frac i 2 	[ \hat{K}_1, \hat{H}_F ]    - \frac {\partial \hat{K}_2} {\partial t}. \nonumber
\end{equation}
We infer 
\begin{equation}
	\hat{H}^{\rm eff}_1(t)  =  \overline{i 	[ \hat{K}_1, \hat{H}_0 ]+ \frac i 2 	[ \hat{K}_1, \hat{H}_F ]  }  
	\;\; \mbox{and} \;\:
	\frac {\partial \hat{K}_2} {\partial t}  =   i 	[ \hat{K}_1, \hat{H}_0 ]+ \frac i 2 	[ \hat{K}_1, \hat{H}_F ]- \hat{H}^{\rm eff}_1.
	\label{eq:K2derivative}
\end{equation}
As a product of a kick operator with a slow Hamiltonian yields a null average, we have $\overline{[ \hat{K}_1(t), \hat{H}_0 ]}=0$ and get the first order expression to the effective Hamiltonian
\begin{equation}
	\hat{H}^{\rm eff}_1(t)= \sum_{m,n} \frac 1 2 \left(\frac {1} {\omega_m}+\frac {1} {\omega_n}  \right)   \left[ \hat{V}_m, \hat{V}_n^{\dagger} \right] e^{  i (\omega_m-\omega_n) t},  
\end{equation}
and the second order expression for the kick operator
\begin{equation}
	\label{eq:K2}
	\hat{K}_2(t)=  \sum_m \frac {1} {i \omega_m^2}  \left( [ \hat{V}_m, \hat{H}_0]  e^{ i \omega_m t } - h.c.  \right) + \sum_{m,n} \frac {1} {2 i \omega_m (\omega_m+\omega_n)} \left( \left[ \hat{V}_m, \hat{V}_n \right] e^{i (\omega_m+\omega_n) t} -  h.c. \right), 
\end{equation}
up to an arbitrary constant operator.

To find $\hat{H}^{\rm eff}_2(t)$, we iterate the very same procedure. Using Eqs.~(\ref{eq:Id1},\ref{eq:Id2}), we find
\begin{eqnarray}
	\hat{H}^{\rm eff}_2(t) = i [\hat{K}_2, \hat{H}] - \frac 1 2 [\hat{K}_1, [\hat{K}_1,\hat{H}]]- \frac {i} {2} [\hat{K}_2, \frac {\partial \hat{K}_1} {\partial t} ] - \frac {i} {2} [ \hat{K}_1, \frac {\partial \hat{K}_2} {\partial t} ] + \frac 1 6 [ \hat{K}_1, [\hat{K}_1,\frac {\partial \hat{K}_1} {\partial t}]] - \frac {\partial \hat{K}_3} {\partial t}
\end{eqnarray}
This equation can be recast using Eqs.~(\ref{eq:K1derivative},\ref{eq:K2derivative}) to express the derivatives of the kick operators $K^{(1,2)}(t)$ in terms of commutators:
\begin{eqnarray}
	\label{eq:K3derivative}
	\hat{H}^{\rm eff}_2(t) =  i [\hat{K}_2, \hat{H}_0]+ \frac i 2 [\hat{K}_2, \hat{H}_F]	+ \frac {i} {2} [\hat{K}_1, \hat{H}^{\rm eff}_1 ] - \frac {1}  {12} [\hat{K}_1, [\hat{K}_1,\hat{H}_F]]   - \frac {\partial \hat{K}_3} {\partial t}
\end{eqnarray}
We find $\overline{[\hat{K}_2, \hat{H}_0]}=\overline{[\hat{K}_1, \hat{H}^{\rm eff}_1]}=0$. As the product of three fast operators with similar Floquet frequencies does not generate any slow harmonics, we have $\overline{[\hat{K}_1, [\hat{K}_1,\hat{H}_F]]}=0. $ The second-order effective Hamiltonian contribution is then given by
\begin{eqnarray}
	\label{eq:H2eff}
	\hat{H}^{\rm eff}_2(t) & = & \frac {i} {2} \overline{[ \hat{K}_2 , \hat{H}_F ]} = \sum_{m,n}   \frac {1} {2 \omega_m^2} \left[ [ \hat{V}_m, \hat{H}_0],  \hat{V}_n^{\dagger} \right] e^{i (\omega_m-\omega_n)t} + h.c.,
\end{eqnarray}
$\hat{K}_3(t)$ can be obtained by an integration of Eq.~\eqref{eq:K3derivative} using the expression~\eqref{eq:H2eff} for $\hat{H}^{\rm eff}_2(t)$. The 3rd-order contribution to the kick operator is derived in Section \ref{sec3} in the specific case $\hat{H}_0= \omega_0 \sigma_z$ and $\hat{V}_m= \Omega \sigma_x.$

\section{DERIVATION OF THE 2ND-ORDER EFFECTIVE MASTER EQUATION}
\label{sec2}
We provide here a detailed derivation the effective Master equation, starting from the following expression obtained in the main text
\begin{equation} 
	\label{eq:StartingPointMasterEquation}
	\frac {\partial \overline{\rho} } {\partial t}=  - i [\hat{H}^{\rm eff}(t), \rho_e]+ \partial_t \overline{\delta \rho^{(2)}(t,\tau)} + \partial_\tau \overline{\delta \rho^{(2)}(t,\tau)}+ \partial_t \overline{\delta \rho^{(3)}(t,\tau)}+O(\Omega \varepsilon^3 ).
\end{equation}
Using the evolution of the instantaneous state [Eq.(1) of the main text], the following expansion of the fast unitary transform in terms of the kick operators
\begin{equation}
	e^{i \hat{K}(t)}= 1-i \hat{K}_1(t) - \frac 1 2 \hat{K}_1(t)^2  - i \hat{K}_2(t) - \frac 1 2 \{ \hat{K}_1(t), \hat{K}_2(t)\} +  \frac i 6 \hat{K}_1(t)^3 - i \hat{K}_3(t) +O\left( \frac {\Omega^4} {\omega^4} \right) \nonumber
\end{equation}
and the properties $\overline{\hat{K}_m(t)}=0$ for $m \geq 1$ and $\overline{\hat{K}_1(t)^{2n+1}}=0$ for $n \geq 0$ mentioned in the main text, 
we obtain the following expressions for $\delta \rho^{(2)}(t,\tau),\delta \rho^{(3)}(t,\tau)$:
\begin{eqnarray}
	\label{eqs:delta2delta3}
	\overline{\delta \rho^{(2)}}(t,\tau) & = &  - \frac 1 2 \{ \overline{\hat{K}_1(t)^2} , \rho_e(\tau) \}	+  \overline{\hat{K}_1(t) \rho_e(\tau) \hat{K}_1(t)} \\
	\overline{\delta \rho^{(3)}}(t,\tau) & = & \overline{\hat{K}_1(t) \rho_e(\tau) \hat{K}_2(t)}+ \overline{\hat{K}_2(t) \rho_e(\tau) \hat{K}_1(t)} - \frac 1 2 \{  \overline{\{ \hat{K}_1(t), \hat{K}_2(t)\}} , \rho_e(\tau) \} \nonumber
\end{eqnarray}
Let us first derive the term $ \partial_t \overline{\delta \rho^{(2)}}(t,\tau).$ Using Eq.~\eqref{eq:K1derivative} and the substitution $\rho_e(\tau)=\overline{\rho}+O(\varepsilon^2)$, we find
\begin{equation}
	\partial_t \overline{\delta \rho^{(2)}}(t,\tau) =- \frac 1 2 \{ \overline{ \{ \hat{H}_F(t),\hat{K}_1(t) \}} ,  \overline{\rho} \}  
	+   \overline{\hat{H}_F(t)  \overline{\rho} \hat{K}_1(t)}+ \overline{ \hat{K}_1(t) \overline{\rho} \hat{H}_F(t)}  +O(\Omega \varepsilon^3) \label{eq:dtrho2}
\end{equation}
Using Eq.~\eqref{eq:K1}, we compute the second contribution of the r.h.s. as
\begin{eqnarray}
	\overline{  \hat{H}_F(t) \overline{\rho}\hat{K}_1(t) } & = & \sum_{m,n} \frac {-1} {i \omega_n} \hat{V}_m \overline{\rho} \hat{V}_n^{\dagger} \overline{e^{i (\omega_m-\omega_n)t}} +
	\sum_{m,n} \frac {1} {i \omega_n} \hat{V}_m^{\dagger} \overline{\rho} \hat{V}_n \overline{e^{i (\omega_n-\omega_m)t}} \nonumber \\
	& = & \sum_{m,n} \frac {1} {i} \left(  \frac { \hat{V}_n^{\dagger} \overline{\rho} \hat{V}_m } {\omega_m} - \frac {\hat{V}_m \overline{\rho} \hat{V}_n^{\dagger}} {\omega_n}  \right)  \overline{e^{i (\omega_m-\omega_n)t}}, \nonumber
\end{eqnarray}
where we have exchanged the indices $m$ and $n$ in the second term. Similarly, we have
\begin{equation}
	\overline{  \hat{K}_1(t) \overline{\rho} \hat{H}_F(t) }= \sum_{m,n} \frac {1} {i} \left( \frac { \hat{V}_m \overline{\rho} \hat{V}_n^{\dagger}} {\omega_m} - \frac {\hat{V}_n^{\dagger} \overline{\rho} \hat{V}_m} {\omega_n}  \right)  \overline{e^{i (\omega_m-\omega_n)t}} \nonumber
\end{equation}
Summing up both contributions , we finally obtain
\begin{eqnarray}
	\overline{  \hat{H}_F(t) \overline{\rho} \hat{K}_1(t) }+	\overline{   \hat{K}_1(t) \overline{\rho} \hat{H}_F(t) } & = & \sum_{m,n} \frac {1} {i} \left( \frac {1} {\omega_m}-\frac {1} {\omega_n} \right)  \overline{e^{i (\omega_m-\omega_n)t}} \left( \hat{V}_m \overline{\rho} \hat{V}_n^{\dagger}+ \hat{V}_n^{\dagger} \overline{\rho} \hat{V}_m\right) \nonumber \\
	& = & \sum_{m<n} \frac {4 \: \overline{\sin (\Delta \omega_{mn} t)}} {\omega_{mn-}} \left( \hat{V}_m \overline{\rho} \hat{V}_n^{\dagger}+ \hat{V}_n^{\dagger} \overline{\rho} \hat{V}_m\right) \label{eq:HFrhoK1av}
\end{eqnarray}
with $\Delta \omega_{nm}= \omega_n-\omega_m$ and $1/\omega_{mn-}=\frac 1 2 (1/\omega_{m}-1/\omega_{n})$. The term $ \{ \overline{ \{ \hat{H}_F(t),\hat{K}_1(t) \}} ,  \overline{\rho} \} $ can be obtained along similar lines. Finally, from Eqs.(\ref{eq:StartingPointMasterEquation},\ref{eq:dtrho2},\ref{eq:HFrhoK1av}), the contribution $\mathcal{L}^{FF}[\overline{\rho}]$ can be expressed as  	$\mathcal{L}^{FF}[\overline{\rho}]=\partial_t \overline{\delta \rho^{(2)}}(t,\tau)$ and
\begin{equation}
	\mathcal{L}^{FF}(\overline{\rho})=\sum_{m<n} \frac {4 \: \overline{\sin (\Delta \omega_{mn} t)}} {\omega_{mn-}} \left( \frac 1 2  \{ \{ \hat{V}_m^{\dagger},\hat{V}_n \} , \rho \} -\hat{V}_m \overline{\rho} \hat{V}_n^{\dagger}- \hat{V}_n^{\dagger} \overline{\rho} \hat{V}_m\right) +O(\Omega \varepsilon^3), \nonumber
\end{equation}
which yields Eq.(4) of the main text.

In the following, we assume that $\overline{e^{\pm i (\omega_m-\omega_n)t}}=e^{\pm i (\omega_m-\omega_n)t},$ and derive the  term ${\cal L}_2^{FSF}[\overline{\rho}] \equiv \partial_\tau \overline{\delta \rho^{(2)}(t,\tau)}+ \partial_t \overline{\delta \rho^{(3)}(t,\tau)}$ - the replacement of $\rho_e(t)$ by $\overline{\rho}(t)$ is valid up to 3rd order corrections. We begin by deriving explicitly $ \partial_\tau \overline{\delta \rho^{(2)}(t,\tau)}$. Special care is needed with respect to the operator ordering. As an example, we find
\begin{eqnarray}
	\frac {\partial }  {\partial \tau}	\left[ \{ \overline{\hat{K}_1(t)^2}, \overline{\rho(\tau)}  \} \right] & = & - i  \{  \overline{\hat{K}_1(t)^2}, [ \hat{H}^{\rm eff}(t), \overline{\rho(\tau)} ]  \} \nonumber \\
	& = & -i [ \hat{H}^{\rm eff}(t), \{  \overline{\hat{K}_1(t)^2}, \overline{\rho(\tau)}  \} ] - i \{ [\overline{ \hat{K}_1(t)^2}, \hat{H}^{\rm eff}(t)] , \overline{\rho
		(\tau)} \}, \nonumber 
\end{eqnarray}
where we have used the generic identity $\hat{A} [\hat{B}, \hat{C}] = [\hat{A},\hat{B}] \hat{C}+ [\hat{B},\hat{A}] \hat{C}$. 
The other terms can be evaluated in a similar manner:
\begin{eqnarray}
	\label{eq:derivative1}
	\frac {\partial \overline{\delta \rho^{(2)}(t,\tau)}} {\partial \tau} & = &  -i [ \hat{H}^{\rm eff}, \overline{\delta \rho^{(2)}(t)} ]  -i \{ [\frac 1 2 \overline{ \hat{K}_1(t)^2}, \hat{H}_0] , \overline{\rho(\tau)} \} 	-i   \overline{[\hat{K}_1(t),\hat{H}_0] \overline{\rho}(\tau) \hat{K}_1(t)} \nonumber \\  & - &  i \overline{  \hat{K}_1(t) \overline{\rho}(\tau) [\hat{K}_1(t),\hat{H}_0]}. 
\end{eqnarray}
We have taken $\hat{H}^{\rm eff}=\hat{H}_0$ in the dissipative terms, which is valid to the considered order.

We now derive the contribution $\partial_t \overline{\delta \rho^{(3)}(t,\tau)}$. Let's  compute  the term $\partial_t \overline{\hat{K}_1(t) \rho_e(\tau) \hat{K}_2(t)}$:
\begin{eqnarray}
	\frac {\partial} {\partial t}  \overline{\hat{K}_1(t) \rho_e(\tau) \hat{K}_2(t)} &=&  \overline{\hat{H}_F(t) \rho_e(\tau) \hat{K}_2(t)}	+ i \overline{\hat{K}_1(t) \rho_e(\tau) [\hat{K}_1(t),\hat{H}_0] } + \frac i 2 \overline{\hat{K}_1(t) \rho_e(\tau) [\hat{K}_1(t),\hat{H}_F(t)] }\nonumber \\  &- & \overline{\hat{K}_1(t) \rho_e(\tau) \hat{H}^{\rm eff}_1(t)}, \nonumber
\end{eqnarray}
where we have used Eqs.~(\ref{eq:K1derivative},\ref{eq:K2derivative}). We find that the two last contributions of the r.h.s. vanish upon averaging, i.e. $\overline{\hat{K}_1 \overline{\rho} \hat{H}^{\rm eff}_1} = \overline{\hat{K}_1 \overline{\rho} [\hat{K}_1,\hat{H}_F] }=0$. Note that the second contribution of the r.h.s. cancels a term from $\frac {\partial \overline{\delta \rho^{(2)}(t,\tau)}} {\partial \tau}$. Other contributions to  $\partial_t \overline{\delta \rho^{(3)}(t,\tau)}$ are obtained along similar lines, and one obtains a one by one cancellation of the dissipative terms in $\partial_{\tau} \overline{\delta \rho^{(2)}(t,\tau)}$. Using Eqs.~\eqref{eq:StartingPointMasterEquation} and Eq.\eqref{eq:derivative1}, we find
\begin{equation}
	\frac {\partial \overline{\rho} } {\partial t}=  - i [\hat{H}_{\rm eff}, \rho_e+ \overline{\delta \rho^{(2)}}]+  	\mathcal{L}^{FF}(\overline{\rho})+ 	\mathcal{L}^{FSF}(\overline{\rho}) +O(\Omega \varepsilon^3). \nonumber
\end{equation}
By writing $\overline{\rho}= \rho_e+ \overline{\delta \rho^{(2)}}+O(\Omega \varepsilon^3)$, the equation above becomes a close equation in $\overline{\rho}$ at the considered order. The contribution coupling the fast and slow quantum dynamics is expressed as
\begin{eqnarray}
	\mathcal{L}^{FSF}[\overline{\rho}]= \overline{\hat{H}_F(t) \overline{\rho}(\tau) \hat{K}_2(t)}+ \overline{\hat{K}_2(t) \overline{\rho}(\tau) \hat{H}_F(t)} - \frac 1 2 \{  \overline{\{ \hat{H}_F(t), \hat{K}_2(t)\}} , \overline{\rho}(\tau) \}. \nonumber
\end{eqnarray}
Let us evaluate one of these terms, for instance $\overline{\hat{H}_F(t) \overline{\rho}(\tau) \hat{K}_2(t)}$. From Eq.~\eqref{eq:K2}, the kick operator $\hat{K}_2(t)$ contains contributions oscillating approximately at the Floquet frequency and at twice the Floquet frequency respectively. The latter does not contribute as it vanishes upon  time-averaging. The considered contribution eventually boils down to
\begin{equation}
	\overline{\hat{H}_F(t) \overline{\rho}(\tau) \hat{K}_2(t)}=   \sum_{m,n} \left(  \frac {1} {i \omega_n^2} \hat{V}_m  \overline{\rho} [\hat{V}_n^{\dagger},\hat{H}_0]  + \frac {1} {i \omega_m^2} \hat{V}_n^{\dagger} \overline{\rho} [\hat{V}_m,\hat{H}_0] \right)  e^{i (\omega_m-\omega_n) t}. \nonumber
\end{equation}
Other terms are derived in a similar manner. Gathering all the contributions, we have 
\begin{eqnarray}
	\mathcal{L}^{FSF}[\overline{\rho}] & = & \frac {i} {2} \sum_{m,n} \left(   \{  \frac {1} {\omega_n^2} \{  \hat{V}_m, [\hat{V}_n^{\dagger},\hat{H}_0]  \}  + \frac {1} {\omega_m^2} \{ \hat{V}_n^{\dagger}, [\hat{V}_m,\hat{H}_0] \}    ,  \overline{\rho} \} \right) e^{i (\omega_m-\omega_n)t}  \nonumber \\ 
	& - & i  \sum_{m,n} \left(  \frac {1} { \omega_n^2} \left(  \hat{V}_m \overline{\rho} [\hat{V}_n^{\dagger},\hat{H}_0] + [\hat{V}_n^{\dagger}, \hat{H}_0] \overline{\rho} \hat{V}_m  \right) +  \frac {1} {\omega_m^2} \left(  \hat{V}_n^{\dagger} \overline{\rho} [\hat{V}_m, \hat{H}_0] + [\hat{V}_m,\hat{H}_0] \overline{\rho} \hat{V}_n^{\dagger} \right) \right)  e^{i (\omega_m-\omega_n) t}. \nonumber
\end{eqnarray}
which can be written more concisely as Eq.(5) of the main text.

\section{HIGHER-ORDER CONTRIBUTIONS TO THE EFFECTIVE MASTER EQUATION}
\label{sec3}

In this section, we work our the method to get the next-order terms to improve the accuracy of the effective quantum Master equation for larger values of the parameters $\varepsilon$ (and larger dissipation strengths), and we provide a few applications related to the examples developed in the main text.

At higher orders, one can no longer substitute $\rho_e$ by $\overline{\rho}$ in the second-order quantum dissipative terms  - this would be equivalent to ignoring terms of similar magnitude as the corrections that we seek to obtain. Consequently, we rely on the relation $\rho_e =\overline{\rho} -  \overline{\delta \rho^{(2)}(t,\tau)}+O(\varepsilon^3)$ within the second-order dissipative contributions. 

Then, the effective Master equation can be written as
\begin{eqnarray} 
\label{eq:StartingPointMasterEquation3rdorder}
\frac {\partial \overline{\rho} } {\partial t} & = &   - i [\hat{H}_{\rm eff}, \rho_e]+ \partial_t \overline{\delta \rho^{(2)}(t,\tau)} + \partial_\tau \overline{\delta \rho^{(2)}(t,\tau)}+ \partial_t \overline{\delta \rho^{(3)}(t,\tau)}  \nonumber \\  & + &  \partial_\tau \overline{\delta \rho^{(3)}(t,\tau)}+ \partial_t \overline{\delta \rho^{(4)}(t,\tau)}
+O(\Omega \varepsilon^4 ).
\end{eqnarray}
where
\begin{eqnarray}
\partial_t \overline{\delta \rho^{(2)}(t,\tau)} & = & \dot{\mathcal{E}}_2[ \overline{\rho} - \mathcal{E}_2[\overline{\rho}]  ] +O( \Omega \varepsilon^5), \nonumber \\
\partial_{\tau} \overline{\delta \rho^{(2)}(t,\tau)} & = & - i \mathcal{E}_2 [ [\hat{H}^{\rm eff}, \overline{\rho} - \mathcal{E}_2[\overline{\rho}   ] ] ]  +O( \Omega  \varepsilon^5), \nonumber \\
\partial_t \overline{\delta \rho^{(3)}(t,\tau)} & = & \dot{\mathcal{E}}_3 [  \overline{\rho} - \mathcal{E}_2[\overline{\rho}] ] +O( \Omega  \varepsilon^5).
\end{eqnarray}
The linear maps $\mathcal{E}_m[\rho]$ are defined in a similar way as in Ref.~\cite{Gamel10}:  $\mathcal{E}_m[\rho]$ is associated to the $m$th-order correction, i.e. these maps are defined by the relation $\mathcal{E}_m[\rho_e]=\overline{\delta \rho^{(m)}}$. From Eqs.~\eqref{eqs:delta2delta3}, we have
\begin{eqnarray}
\mathcal{E}_2[\rho]  & = & - \frac 1 2 \{ \overline{\hat{K}_1(t)^2}, \rho \}+ \overline{\hat{K}_1(t) \rho \hat{K}_1(t)}  \\
\mathcal{E}_3[\rho]  & = &	 \overline{\hat{K}_1(t) \rho \hat{K}_2(t)}+ \overline{\hat{K}_2(t) \rho \hat{K}_1(t)} - \frac 1 2 \{  \overline{\{ \hat{K}_1(t), \hat{K}_2(t)\}} , \rho \} 
\end{eqnarray}
The time dependence of the linear maps $\mathcal{E}_m $ comes from the operators $\hat{K}_m(t)$ - $\rho$ is used as a simple variable on which the map is applied.
We have also used the fact that the term $  \dot{\mathcal{E}}_2[ \overline{\rho}  ] \equiv \mathcal{L}^{(c)}[\overline{\rho}]$ is already of second-order in $\varepsilon$.

\subsection{3rd and 4th-order contributions in the Phase Noise configuration ($\hat{H}_0 \propto \sigma_z$ and $\hat{H}_F \propto \sigma_z$)}

In the specific case of phase noise, one can easily obtain the effective equation up to the 5th order. Indeed, one has $\hat{K}_m=0$ for $m \geq 2$, so the total kick operator is simply $\hat{K}(t)=\hat{K}_1(t)$. Then, the different non-unitary terms arise from an expansion of the unitary operators
$e^{\pm i \hat{K}_1(t)}$.  Odd-power of the kick operator $\hat{K}_1$ disappear upon time averaging, so that $\mathcal{E}_1=\mathcal{E}_3=\mathcal{E}_5=0$,
and	the 4th-order equation is given by
\begin{equation}
\label{eq:higherordereffectiveequationgeneral}
\dot{\overline{\rho}} = - i  [ \hat{H}_{\rm eff}, \overline{\rho}  ] + \dot{\mathcal{E}}_2[\overline{\rho}]  +\dot{\mathcal{E}}_4[\overline{\rho}]- \dot{\mathcal{E}}_2[\mathcal{E}_2[\overline{\rho}]] +O(\Omega \varepsilon^5)
\end{equation}
As $\mathcal{E}_5=0$, the next-order terms correspond to $\dot{\mathcal{E}}_6[\rho]$ and are thus of 5th-order. We have used the fact that for the present case for which the operators $\hat{K}_1$ and  $\hat{H}^{\rm eff}$ are both proportional to $\sigma_z$, one has $  [\hat{H}^{\rm eff},  \mathcal{E}_2[\overline{\rho}   ] ] =0$. By expanding $e^{i \hat{K}_1(t)}$ in the expression of the evolved quantum state (Eq.(1) of the main text), one finds the 4th-order expansion (we note $\mathcal{E}_4[\rho]\equiv \mathcal{E}_4^{(c)}[\rho]$ this  ``commutative'' contribution associated to the kick operators $\hat{K}_1(t)$ alone)
\begin{eqnarray}
\label{eq:E4commuting}
\mathcal{E}_4^{(c)}[\rho] & = & \frac {1} {24}  \{ \overline{\hat{K}_1^{4}}   , \rho   \} - \frac 1 6 \overline{\hat{K}_1^3 \rho \hat{K}_1} - \frac 1 6 \overline{\hat{K}_1 \rho \hat{K}_1^3} 
+ \frac 1 4 \overline{\hat{K}_1^{2} \rho \hat{K}_1^{2} }	
\end{eqnarray}
We then express the corresponding time derivatives
\begin{eqnarray}
\dot{\mathcal{E}}_4^{(c)}[\rho] & = & \frac 1 6 \{ \overline{\hat{H}_F \hat{K}_1^{3}}, \rho \}  - \frac 1 2 \overline{ \hat{H}_F \hat{K}_1^{2} \rho \hat{K}_1} 
- \frac 1 6 \overline{\hat{K}_1^{3} \rho \hat{H}_F}  - \frac 1 2 \overline{  \hat{K}_1 \rho   \hat{H}_F \hat{K}_1^{2}}  - \frac 1 6 \overline{\hat{H}_F \rho \hat{K}_1^{3} }   \nonumber \\  &+& \frac 1 2 \overline{\hat{K}_1^{2} \rho \hat{K}_1 \hat{H}_F } + \frac 1 2 \overline{\hat{H}_F  \hat{K}_1 \rho \hat{K}_1^{2} }, \nonumber \\ 
\dot{\mathcal{E}}_2[ \mathcal{E}_2 [\rho]] & = & -  \{ \overline{\hat{H}_F \hat{K}_1}, - \frac 1 2 \{ \overline{\hat{K}_1^2}, \rho \} + \overline{\hat{K}_1 \rho \hat{K}_1} \:  \}
+ \overline{\hat{H}_F \left(  - \frac 1 2 \{ \overline{\hat{K}_1^2}, \rho \} + \overline{\hat{K}_1 \rho \hat{K}_1} \right) \hat{K}_1}  \nonumber \\ &+& \overline{\hat{K}_1 \left(  - \frac 1 2 \{ \overline{\hat{K}_1^2}, \rho \} + \overline{\hat{K}_1 \rho \hat{K}_1} \right) \hat{H}_F} ,\nonumber
\end{eqnarray}
where we have used the commutation relation $[\hat{K}_1,\hat{H}_F]=0$. One can drastically simplify these expressions by writing $\hat{K}_1(t)= F(t) \sigma_z$, $\hat{H}_F(t)= f(t) \sigma_z$, and using the identity $\sigma_z^2= 1_{2 \times 2}$:
\begin{eqnarray}
\dot{\mathcal{E}}_4^{(c)}[\rho] & = & \frac 4 3 \: \overline{f(t) F(t)^3}\left( \rho- \sigma_z \rho \sigma_z \right), \label{eq:E4simplifie}\\
\dot{\mathcal{E}}_2[ \mathcal{E}_2 [\rho]] & = & 2 \: \overline{f(t) F(t)} \: \: \overline{F(t)^2} \left( \rho- \sigma_z \rho \sigma_z \right). \label{eq:E2E2simplifie}
\end{eqnarray}
We take as in the main text  $\hat{V}_m=\Omega \sigma_z$. The functions respectively associated to the Floquet Hamiltonian and kick operator~\eqref{eq:K1} are given by  $f(t)= \Omega \sum_{m, \varepsilon_m} e^{i \varepsilon_m \omega_m t} $ and $F(t)= \Omega \sum_{m, \varepsilon_m} \frac {\varepsilon_m} {i \omega_m} e^{i \varepsilon_m \omega_m t} $, where for each label $m$ the sum is extended over all the Floquet frequencies, and the label $\varepsilon_m$ takes the two values $\{-1, 1\}$.  

From previous results, the second-order time-averaged functions read
\begin{eqnarray}
\overline{f(t) F(t)} = 	 \Omega^2 \sum_{m,n} \left(     \frac { e^{i  (\omega_n-\omega_m) t}-e^{i  (\omega_m-\omega_n) t}}  {i \omega_n}  \right) \;\;\mbox{and} \;\;\overline{F(t)^2} = 2   \Omega^2 \sum_{m,n} \frac {1} { \omega_m  \omega_n }    e^{i  (\omega_m-\omega_n) t}. \nonumber
\end{eqnarray}
Let us evaluate the fourth-order time-averaged function 
\begin{equation}
\overline{f(t) F(t)^3} = \Omega^4 \sum_{m,\varepsilon_m} \sum_{n,\varepsilon_m} \sum_{p,\varepsilon_p} \sum_{q,\varepsilon_q}  e^{i \varepsilon_m \omega_m t} \frac {\varepsilon_n e^{i \varepsilon_n \omega_n t}} {i \omega_n} \frac {\varepsilon_p e^{i \varepsilon_p \omega_p t}} {i \omega_p} \frac {\varepsilon_q e^{i \varepsilon_q \omega_q t}} {i \omega_q} \delta_{\varepsilon_m+\varepsilon_n+\varepsilon_p+\varepsilon_q,0} \nonumber 
\end{equation}
The Kronecker symbol $\delta_{\varepsilon_m+\varepsilon_n+\varepsilon_p+\varepsilon_q,0}$ accounts for the time-averaging and retains only the slow-rotating contributions such that $\varepsilon_m+\varepsilon_n+\varepsilon_p+\varepsilon_q=0$. For a given $\varepsilon_m=\pm 1$, there are only $3$ sets $\{\varepsilon_m,\varepsilon_p,\varepsilon_q \}$ in $\{-1,1 \}^3$ that yield  $\varepsilon_m+\varepsilon_n+\varepsilon_p+\varepsilon_q=0$, so that
\begin{eqnarray}
\overline{f(t) F(t)^3} & = & 3 \: \Omega^4  \sum_{m,n,p,q} \left(  e^{i  \omega_m t}  \frac {e^{i  \omega_n t}} {i \omega_n}  \frac { e^{- i \omega_p t}} {i \omega_p} \frac { e^{- i  \omega_q t}} {i \omega_q} -  e^{-i  \omega_m t}  \frac {e^{i  \omega_n t}} {i \omega_n}  \frac { e^{- i \omega_p t}} {i \omega_p} \frac { e^{ i  \omega_q t}} {i \omega_q} \right) \nonumber
\end{eqnarray}
which can be rewritten as
\begin{eqnarray}
\overline{f(t) F(t)^3} & = & 3 \: \Omega^4  \sum_{n,p} \left(     \frac {e^{i  \omega_n t}} {i \omega_n}  \frac { e^{- i \omega_p t}} {i \omega_p}  \right)
\sum_{m,q} \left( e^{i  \omega_m t} \frac { e^{- i  \omega_q t}} {i \omega_q} -e^{-i  \omega_m t} \frac { e^{ i  \omega_q t}} {i \omega_q} \right) \nonumber \\
& = & 3 \:  \frac {\overline{F(t)^2}} {2}  \: \: \overline{f(t) F(t)} \nonumber
\end{eqnarray}
From Eqs.(\ref{eq:E4simplifie},\ref{eq:E2E2simplifie}), one finds $\dot{\mathcal{E}}_4^{(c)}[\rho]-	\dot{\mathcal{E}}_2[ \mathcal{E}_2 [\rho]]=0$, i.e. the two 3rd-order terms of the equation cancel each other. As there are no 4th-order terms, the equation presented in the main text is accurate to the 5th order. \\


\subsection{3rd order contribution to the effective Equation in the configuration $\hat{H}_0  \propto \sigma_z$ and $\hat{H}_F \propto \sigma_x$}

From Eq.(1) of the main text and expansion of the exponential of kick operators, one obtains the  4th-order contribution to the density matrix $\overline{\delta \rho^{(4)}}=\mathcal{E}_4^{(c)}[\overline{\rho}]+ \mathcal{E}_4^{(nc)}[\overline{\rho}] $ where 
\begin{equation}
\mathcal{E}_4^{(nc)}[\overline{\rho}]= - \frac 1 2 \{ \overline{\{ \hat{K}_1, \hat{K}_3 \}} , \overline{\rho} \} + \overline{\hat{K}_1 \overline{\rho} \hat{K}_3}+ \overline{\hat{K}_3 \overline{\rho} \hat{K}_1} + \overline{\hat{K}_2 \overline{\rho} \hat{K}_2}. \nonumber
\end{equation}
This contribution arises from the non-commutation of the Floquet and constant Hamiltonians. As seen previously, the ``commutating'' contribution  $\dot{\mathcal{E}}_4^{(c)}[\overline{\rho}]$ to the equation of motion is canceled by the term $-\dot{\mathcal{E}}_2[\mathcal{E}_2[\overline{\rho}]]$ (the calculation performed above relies on $[\hat{H}_F,\hat{K}_1]=0$ and thus still holds here).

Using Eq.~\eqref{eq:StartingPointMasterEquation}, the additional 3rd-order terms to the effective Master equation correspond to:
\begin{equation}
\mathcal{L}^{(3)}[\overline{\rho}]= i [\hat{H}^{\rm eff},\overline{\delta \rho^{(3)}}]+	\partial_{\tau} \overline{\delta \rho_3}(t,\tau) +	\dot{\mathcal{E}}_4^{(nc)}[\overline{\rho}].  \nonumber
\end{equation}
The first term arises from the identity $\rho_e=\overline{\rho}-\overline{\delta \rho^{(2)}}-\overline{\delta \rho^{(3)}}+O(\varepsilon^4)$  in the commutator $[\hat{H}^{\rm eff},\rho_e]$ of Eq.~\eqref{eq:higherordereffectiveequationgeneral}.

Before evaluating the 3rd-order contributions, we give the expression for the kick operators in this case. With $\hat{H}_0 = \omega_0  \sigma_z$ and $\hat{V}_m = \Omega \sigma_x$, the two leading kick operators read   $\hat{K}_1(t)= \Omega \sum_m \frac {e^{i \omega_m t}}
{i \omega_m } \sigma_x +h.c. $ and $\hat{K}_2(t)= - 2 \omega_0 \Omega  \sum_m \frac {e^{ i \omega_m t }} { \omega_m^2}     \sigma_y + h.c. $. From Eqs.~(\ref{eq:K3derivative},\ref{eq:H2eff}), one obtains the 3rd-order kick operator:
\begin{equation}
\label{eq:K3}
\hat{K}_3(t)= \sum_m \frac {4 \omega_0^2 \Omega} {i \omega_m^3} \left(e^{i \omega_m t }-h.c. \right) \sigma_x  - 2 \sum_{mn}  \frac {\omega_0 \Omega^2} {i \omega_m^2 (\omega_m+\omega_n)} \left( e^{i(\omega_m+\omega_n)t} -h.c. \right) \sigma_z.
\end{equation}

The 3rd-order correction derived in the main text reads
\begin{equation} 
\overline{\delta \rho_3}(t,\tau) =  \overline{\hat{K}_1(t) \rho_e(\tau) \hat{K}_2(t)}+ \overline{\hat{K}_2(t) \rho_e(\tau) \hat{K}_1(t)} \nonumber
\end{equation}
as $\{ \hat{K}_1(t),\hat{K}_2(t)  \}=0$ in this specific case.

Let us first evaluate
\begin{equation}
\label{eq:dtaudelta3}
\partial_{\tau} \overline{\delta \rho^{(3)}}(t,\tau) = - i [\hat{H}_{\rm eff},\overline{\delta \rho^{(3)}}] - i  \overline{[\hat{K}_1,\hat{H}_0 ] \overline{\rho} \hat{K}_2} - i  \overline{\hat{K}_2 \overline{\rho} [\hat{K}_1,\hat{H}_0 ] }  -i  \overline{\hat{K}_1 \overline{\rho} [\hat{K}_2,\hat{H}_0 ] }  -i  \overline{[\hat{K}_2,\hat{H}_0] \overline{\rho} \hat{K}_1 }  
\end{equation}
where at this order it is valid to use the equality $\rho_e=\overline{\rho}$ and $\hat{H}_{\rm eff} \equiv \hat{H}_0$ in the dissipative terms. The first term of the r.h.s  yields the correct unitary dynamics.
We now focus on non-unitary terms:
\begin{eqnarray}
\label{eq:E4derivative}
\dot{\mathcal{E}}_4^{(nc)}[\overline{\rho}] & = &  - \frac 1 2 \{ \overline{\{ \hat{K}_1, \frac {d \hat{K}_3} {dt} \}} , \rho \} - \frac 1 2 \{ \overline{\{ \frac {d \hat{K}_1} {dt}, \hat{K}_3 \}} , \rho \} +  \overline{\frac {d \hat{K}_1} {dt} \rho \hat{K}_3}+\overline{\hat{K}_1 \rho \frac {d \hat{K}_3} {dt}} +\overline{\frac {d \hat{K}_3} {dt} \rho \hat{K}_1}+\overline{\hat{K}_3 \rho \frac {d \hat{K}_1} {dt}} \nonumber \\ &+&  \overline{\frac {d \hat{K}_2} {dt} \rho \hat{K}_2}+  \overline{\hat{K}_2 \rho \frac {d \hat{K}_2} {dt}}  
\end{eqnarray}
From  Eq.~\eqref{eq:K2derivative}, we infer
$	\frac {d \hat{K}_2} {dt} = i [\hat{K}_1, \hat{H}_0]$
as $[\hat{K}_1,\hat{H}_F]=0$ for the considered Floquet Hamiltonian. Hence, the terms $\overline{\hat{K}_2 \rho \frac {d \hat{K}_2} {dt}}$ and $\overline{\frac {d \hat{K}_2} {dt} \rho \hat{K}_2}$ simply cancel part of the non-unitary contributions of Eq.\eqref{eq:dtaudelta3}. Thanks to the relation
\begin{equation}
\frac {d \hat{K}_3} {d t} = i [ \hat{K}_2, \hat{H}_0]+ \frac i 2 [\hat{K}_2, \hat{H}_F]-\hat{H}_{\rm eff}^{(2)}, \nonumber
\end{equation}
one can express the contributions of Eq.~\eqref{eq:E4derivative} as
\begin{equation}
\overline{\frac {d \hat{K}_3} {dt} \rho \hat{K}_1}=i \overline{[ \hat{K}_2, \hat{H}_0] \rho \hat{K}_1 }+ \frac {i} {2} \overline{ [\hat{K}_2, \hat{H}_F] \rho \hat{K}_1}-  \overline{ \hat{H}_{\rm eff}^{(2)} \rho \hat{K}_1}. \nonumber
\end{equation}
The first member of the r.h.s. cancels non-unitary contributions of Eq.\eqref{eq:dtaudelta3}. The second and third terms of the r.h.s. yield a null time average. We eventually get the simple relation
\begin{equation}
\mathcal{L}^{(3)}[\overline{\rho}]=  - \frac 1 2 \{ \overline{\{ \hat{H}_F, \hat{K}_3 \}} , \rho \} +  \overline{\hat{H}_F \rho \hat{K}_3} +  \overline{\hat{K}_3 \rho \hat{H}_F} ,\nonumber
\end{equation}
where we have used Eq.\eqref{eq:K1derivative}. These contributions can be evaluated thanks to Eq.\eqref{eq:K3}. Only the terms rotating at a single Floquet frequency contribute to the time-averaging. One obtains
$
\mathcal{L}^{(3)}[\overline{\rho}] = h(t) \left( \overline{\rho} -  \sigma_x \overline{\rho}  \sigma_x \right)
$
with 
\begin{equation}
h(t) = 8\omega_0^2 \Omega^2 \sum_{m,\varepsilon_m, n, \varepsilon_n}  \frac {\varepsilon_m e^{i (\varepsilon_m \omega_m+\varepsilon_n \omega_n)t}} {i \omega_m^3} \delta_{\varepsilon_m+\varepsilon_n,0}. \nonumber
\end{equation}
In the bichromatic case, we find the additional 3rd-order non-unitary contribution
\begin{equation}
\mathcal{L}^{(3)}[\overline{\rho}] = 16 \omega_0^2 \Omega^2 \left( \frac {1} {\omega_1^3}- \frac {1} {\omega_2^3} \right) \sin (\omega_2-\omega_1)t \left( \overline{\rho} -  \sigma_x \overline{\rho}  \sigma_x \right).  \label{eq:L3final}
\end{equation}
This ``third-order'' correction is actually of 4th-order in $\varepsilon$.  A complete treatment of the 4th-order corrections should also include the following contributions in the r.h.s of the effective equation: $ i \mathcal{E}_2 [ \: [\hat{H}_0,  \mathcal{E}_2[\overline{\rho}   ] ] \: ]  -\dot{\mathcal{E}}_3 [  \mathcal{E}_2[\overline{\rho}] ]  
+ \partial_\tau \overline{\delta \rho^{(4)}(t,\tau)}+ \partial_t \overline{\delta \rho^{(5)}(t,\tau)}$ \:. The derivation of these 4th-order terms is a long but straightforward calculation, beyond the scope of this article.

Figure \ref{fig:Supplementary} represents the instantaneous (solid gray line) and time-convoluted density matrix population (solid black line) as a function of time, confronted to the predictions of the second (black dotted line) and third-order (dashed black line) effective Master equations.The latter corresponds to the addition of the contribution $\mathcal{L}_3[\rho]$~\eqref{eq:L3final} to the r.h.s of the 2nd-order effective Master equation [Eq.(3) of the main text].

We have used an initial density matrix $\rho_0=|e\rangle \langle e|$ corresponding to a pure eigenstate of the constant Hamiltonian $\hat{H}_0$. This initial state is also. an eigenstate of the effective Hamiltonian $\hat{H}^{\rm eff}(t)$ in the presence of the Floquet driving. Hence, the unitary part of the quantum dynamics leaves the initial density matrix invariant, and the observed time-dependence in the population comes exclusively from the non-unitary contributions. As in Figs.1,2,3 of the main text, the initial condition for the considered effective equations is obtained from a convolution with the instantaneous solution $\rho(t)$ as $\overline{\rho}_0=\int_{- \infty}^{+ \infty} dt f(-t) \rho(t)$. Figure \ref{fig:Supplementary} reveals that the higher-order correction $\mathcal{L}^{(3)}[\overline{\rho}]$ to the effective Master equation considerably enhances its accuracy in the prediction of the time-coarse grained dynamics.


\begin{figure}[htbp]
\centering
\includegraphics[width=12 cm]{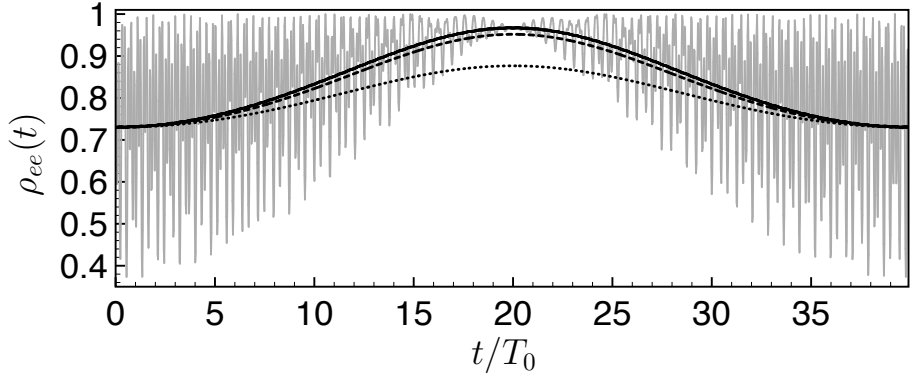}  
\caption{Higher-order effective Master equation vs full quantum evolution: Instantaneous density matrix population $\rho_{ee}(t)$ (solid gray line) as a function of time (in arbitrary unit $T_0$). Time-convoluted population $\overline{\rho}_{{\rm ex}}(t)$ (solid black line) obtained from the exact unitary evolution. The time-coarse grained density matrix population $\overline{\rho}_{ee}(t)$  obtained from the second-order   (dotted black line) and third-order (dashed black line) effective Master equation. Parameters: initial density matrix $\rho_0=| e \rangle \langle e|$, constant 
	and Floquet Hamiltonians $\hat{H}_0=\omega_0 \sigma_z$, $\hat{H}_F(t)= \Omega \left(\cos( \omega_1 t)+\cos( \omega_2 t) \right) \sigma_x.$ Results obtained for the  frequencies $\omega_0=0.5 \times (2 \pi)/ T_0$, $\Omega=3.5/T_0$, and $\varepsilon \simeq 0.18$. Same frequencies $\omega_1$,$\omega_2$,$\omega_c$ as in Figs.2-3 of the main text.}
\label{fig:Supplementary}
\end{figure}


\end{widetext}
	

\begin{thebibliography}{10}
\bibitem{Footnote1}	This distinction is here a mere convenience in order to establish the effective Master equation -- in the end we set $\tau=t$.
\bibitem{Footnote2}	For Lindblad terms associated to $\hat{L}_m=\hat{L}_m^{\dagger}$, e.g. the phase noise Linbladian, this condition is trivially fulfilled.
\end{thebibliography}
\end{document}